\documentclass[useAMS,usenatbib,iop,apj,tighten,11pt,a4paper]{emulateapj}

\usepackage{graphicx,color}
\usepackage{amsmath}
\usepackage{amssymb}
\usepackage{bm}

\def\aap{A\&A}
\def\apjl{ApJ}

\def\apjs{ApJS}

\renewcommand{\vec}[1]{{{\mbox{\boldmath $#1$}}}}

\newcommand{\zhat}{\hat{z}}
\newcommand{\phihat}{\hat{\phi}}

\newcommand{\del}{\partial}

\newcommand{\bfDel}{\bm{\nabla}}

\newcommand{\bmDeltilde}{\widetilde{\bm{\nabla}}}

\newcommand{\bmOmega}{\bm{\Omega}}

\newcommand{\alp}{\alpha}

\newcommand{\Emf}{\bm{\mathcal{E}}}

\newcommand{\Flux}{\bm{\mathcal{F}}}

\newcommand{\bfu}{\bm{u}}
\newcommand{\bfb}{\bm{b}}

\newcommand{\mean}[1]{\overline{#1}}
\newcommand{\meanv}[1]{\overline{\bm{#1}}}

\newcommand{\corr}{_\mathrm{c}}						
\newcommand{\eq}{_\mathrm{eq}}						
\newcommand{\f}{_\mathrm{0}}					   	
\newcommand{\kin}{_\mathrm{k}}			   		
\newcommand{\magn}{_\mathrm{m}}			   		
\newcommand{\cray}{_\mathrm{cr}}			   		
\newcommand{\crit}{_\mathrm{c}}			   		
\newcommand{\diff}{_\mathrm{d}}			   		

\newcommand{\cro}{\times}

\newcommand{\mbr}{\mean{B}_r}
\newcommand{\mbp}{\mean{B}_\phi}
\newcommand{\mbz}{\mean{B}_z}

\newcommand{\muz}{\mean{U}_z}

\newcommand{\rtilde}{\widetilde{r}}

\newcommand{\Btilde}{\widetilde{B}}
\newcommand{\bmBtilde}{\widetilde{\bm{B}}}

\newcommand{\ztilde}{\widetilde{z}}

\newcommand{\Alf}{_\mathrm{A}}

  \newcommand{\kms}{\,{\rm km\,s^{-1}}}
  \newcommand{\kmskpc}{\,{\rm km\,s^{-1}\,kpc^{-1}}}
  \newcommand{\skpckm}{\,{\rm s\,kpc\,km^{-1}}}

  \newcommand{\cmcms}{\,{\rm cm^2\,s^{-1}}}

  \newcommand{\kpc}{\,{\rm kpc}}

  \newcommand{\Myr}{\,{\rm Myr}}
  
  \newcommand{\Gyr}{\,{\rm Gyr}}

  \newcommand{\mkG}{\,\mu{\rm G}}

\definecolor{webgreen}{rgb}{0,.5,0}
\definecolor{webbrown}{rgb}{.6,0,0}
\definecolor{purple}{rgb}{0.5,0,.5}

\begin{document}

\title{Nonlinear Galactic Dynamos and The Magnetic Pitch Angle}
\author{Luke Chamandy and A.~Russ Taylor}
\affil{Astronomy Department, University of Cape Town, Rondebosch 7701, Republic of South Africa}
\affil{Department of Physics, University of the Western Cape, Belleville 7535, Republic of South Africa}
\email{luke@ast.uct.ac.za}
\email{russ@ast.uct.ac.za}

\begin{abstract}
Pitch angles $p$ of the large-scale magnetic fields $\meanv{B}$ of spiral galaxies
have previously been inferred from observations 
to be systematically larger in magnitude than predicted by standard mean-field dynamo theory.
This discrepancy is more pronounced if dynamo growth has saturated,
which is reasonable to assume given that such fields are generally inferred 
to be close to energy equipartition with the interstellar turbulence.
This ``pitch angle problem" is explored using local numerical mean-field dynamo solutions as well as asymptotic analytical solutions.
It is first shown that solutions in the saturated or kinematic regimes depend on only five dynamo parameters,
two of which are tightly constrained by observations of galaxy rotation curves.
The remaining 3-dimensional (dimensionless) parameter space can be constrained to some extent using theoretical arguments.
Predicted values of $|p|$ can be as large as $\sim40^\circ$,
which is similar to the largest values inferred from observations,
but only for a small and non-standard region of parameter space.
We argue, based on independent evidence, that such non-standard parameter values are plausible.
However, these values are located toward the boundary of the allowed parameter space,
suggesting that additional physical effects may need to be incorporated.
We therefore suggest possible directions for extending the basic model considered.
\end{abstract}


\section{Introduction}
The pitch angle $p$ of the regular (i.e. large-scale) magnetic field 
may be the most important observable with which to test galactic dynamo theory.
The quantity $p$ measures how tightly wound is the large-scale field,
and is given by $\tan p=\mbr/\mbp$ where $r$ is the galactocentric radius and $\phi$ is the azimuthal coordinate,
oriented along the direction of the rotational velocity.
By convention, $p$ is defined to be between $-90^\circ$ and $90^\circ$.
A determination of $p$ from polarized radio emission generally involves some amount of inference,
but $p$ is closer to a direct observable than the strength of the mean-field $\mean{B}$.
Note that a larger magnitude of $p$ implies more open magnetic field lines, 
$p=0$ corresponds to a circular field, and $p<0$ corresponds to a trailing magnetic spiral.
$p$ is generally observed to have a negative sign, in agreement with the predictions of mean-field dynamo theory
for a differentially rotating galactic disk, with angular velocity decreasing outward.

Observed magnetic pitch angles have magnitudes larger than those predicted by kinematic mean-field dynamo theory.
In the data set of \citet{Vaneck+15}, the mean value of $p$ is $-25^\circ$,
and $-p$ may be as small as $8^\circ$ or as large as $48^\circ$ (excluding uncertainties),
whereas kinematic theory generally predicts $-p<20^\circ$.
Moreover, this mismatch between theory and observation worsens in the nonlinear regime,
when the field approaches energy equipartition with turbulence \citep{Elstner05,Vaneck+15},
and is predicted to then have a reduced value of $-p$.
As local galaxies have large-scale fields near equipartition, 
it is reasonable to infer that they have already reached the saturated state.
This adds up to the existence of a ``pitch angle problem."
This problem has been addressed in the past by making the disc thinner,
the shear smaller, assuming that the kinetic $\alpha$ effect becomes enhanced in spiral arms \citep{Chamandy+14a},
incorporating mean radial flows \citep{Moss+00}, or invoking spiral shocks \citep{Vaneck+15}.
Although these proposed remedies may all play a role, 
none is likely to be sufficient on its own;
a more natural and universal solution to the problem would be preferred.

The goal of the present paper is a systematic study of the magnetic pitch angles predicted by standard mean-field galactic dynamo theory.
Unlike most previous works, we make no \textit{a priori} judgements about the dynamo parameters;
rather, we focus on mapping out the available parameter space and exploring the solutions 
that may lead to values of $p$ that are in good agreement with observations.
In Section~\ref{sec:theory} we present the dynamo solutions used and summarize the underlying theory.
We then discuss important constraints on the model
stemming from theoretical considerations in Section~\ref{sec:theor_constraints}.
This is followed by our main results in Section~\ref{sec:results},
namely the predictions for the magnetic pitch angle from numerical as well as analytical solutions
over a wide range of parameter values.
In Section~\ref{sec:discussion} we assess, citing independent evidence, the plausibility of various parameter values,
and in Section~\ref{sec:extensions} we offer several ideas for extending the basic model to include additional physical ingredients
that may have important effects on the magnetic pitch angle.
Finally, we summarize and present our conclusions in Section~\ref{sec:conclusions}.

\section{Theoretical background}
\label{sec:theory}
The evolution of the mean magnetic field is governed by the mean-field induction equation
\begin{equation}
  \label{meaninduction}
  \del\meanv{B}/\del t=\bfDel\cro\left(\meanv{U}\cro\meanv{B}+\Emf-\eta\bfDel\cro\meanv{B}\right),
\end{equation}
where overbars represent mean (large-scale) quantities. 
For statistically isotropic turbulence, 
the first order smoothing closure approximation (FOSA) gives the mean electromotive force as
\begin{equation}
  \label{emf}
  \Emf=\alpha\meanv{B}-\beta\bfDel\cro\meanv{B},
\end{equation}
where $\alpha$ and $\beta$ depend on the statistical properties 
of the random (small-scale) velocity $\bfu$ and random magnetic field $\bfb$.
Under the same circumstances, the turbulent diffusivity is given by $\beta=\tfrac{1}{3}\tau\corr u^2$,
where $u\equiv\sqrt{\mean{\bfu^2}}$ and $\tau\corr$ is the correlation time of the random flow.
Combining equations \eqref{meaninduction} and \eqref{emf}, and neglecting $\eta$ compared with $\beta$, 
we obtain the standard dynamo equation
\begin{equation}
  \label{dynamo}
  \del\meanv{B}/\del t=\bfDel\cro\left(\meanv{U}\cro\meanv{B}+\alpha\meanv{B}-\beta\bfDel\cro\meanv{B}\right).
\end{equation}

We adopt the dynamical quenching formalism, 
whereby the quenching of the mean-field dynamo can be understood to arise from 
the additional requirement of magnetic helicity balance.
In this paradigm,
the backreaction of the Lorentz force onto the dynamo is modeled by including magnetic as well as kinetic terms in $\alpha$
\citep{Pouquet+76, Kleeorin+Ruzmaikin82, Gruzinov+Diamond94, Blackman+Field00, Blackman+Brandenburg02, Radler+03, Brandenburg+Subramanian05a}.
Thus, $\alpha=\alpha\kin+\alpha\magn$, with
\begin{equation}
  \label{alpha}
  \alpha\kin=-\tfrac{1}{3}\tau\corr\mean{\bfu\cdot\bfDel\cro\bfu}, \qquad \alpha\magn=\tfrac{1}{3}\tau\corr\mean{\bfu\Alf\cdot\bfDel\cro\bfu\Alf},
\end{equation}
where $\bfu\Alf=\bfb/\sqrt{4\pi\rho}$ is the Alfven velocity associated with the random magnetic field and $\rho$ is the gas density.
Because $\alpha\magn$ is closely related to the mean small-scale magnetic helicity density,
magnetic helicity balance can be invoked to derive an evolution equation for $\alpha\magn$ \citep{Shukurov+06},
\begin{equation}
  \label{dalpha_mdt}
  \del\alpha\magn/\del t=-(2\beta/l^2)(\Emf\cdot\meanv{B}/B\eq^2)-\bfDel\cdot\Flux,
\end{equation}
where $B\eq\equiv u\sqrt{4\pi\rho}$ is the equipartition field strength, $l$ is the correlation scale of the random flow,
and we have neglected an Ohmic dissipation term because it is negligible compared to the term involving the flux density $\Flux$ of $\alpha\magn$.
The latter is given by a sum of advective \citep{Subramanian+Brandenburg06} and diffusive \citep*{Brandenburg+09} terms
\begin{equation}
  \label{flux}
  \Flux=\meanv{U}\alpha\magn -\kappa\bfDel\alpha\magn,
\end{equation}
with $\kappa$ the turbulent diffusivity of $\alpha\magn$.
We note that other terms may be important in equation \eqref{emf} \citep{Brandenburg+Subramanian05a} 
or in equation \eqref{flux} \citep{Subramanian+Brandenburg06,Vishniac12b,Vishniac+Shapovalov14,Ebrahimi+Bhattacharjee14}, 
but they are less certain and so are left for consideration in a future study.

We assume axisymmetry, adopt cylindrical $(r, \phi, z)$ coordinates, and make the local `slab' approximation.
That is we neglect radial as compared to vertical derivatives, e.g. $\del\mbz/\del r\ll\del\mbr/\del z$,
and assume $\mbz^2\ll\mbr^2$.
Generally, this is justified if the local galactocentric radius 
greatly exceeds the disk half-thickness $r\gg h$ \citep{Ruzmaikin+88}.
The resulting equations are solved numerically to obtain $\mbr(t,z)$, $\mbp(t,z)$ and $\alpha\magn(t,z)$.
We adopt vacuum boundary conditions $\mbr=\mbp=0$ as well as $\del^2\alp\magn/\del z^2=0$ at $z=\pm h$;
we also set $\alpha\magn=0$ at $t=0$, while $\mbr$ and $\mbp$ are set to small initial values $\ll B\eq$ 
(the solution for the mean field in the saturated regime is independent of the seed field so long as it is sufficiently small).
For numerical solutions, following \citet{Chamandy+14b} we adopt
\begin{equation*}
  B\eq=B\f\exp\left(-\frac{z^2}{2h^2}\right),
\end{equation*}
where $B\f$ is the strength of the equipartition field at the midplane.
Admittedly, exponential profiles may provide better fits to more observations 
or direct numerical simulations than Gaussian profiles
\citep[see, e.g.][where both exponentials and Gaussians are used to model synchrotron emission in edge-on galaxies]{Dumke+95}.
In any case, the solution for $\meanv{B}$ turns out to be remarkably insensitive to the profile chosen,
with vertically averaged $p$ in the saturated state differing
by $<1\%$ if $B\eq=B\f\exp(-|z|/h)$ or even $B\eq=B\f$ is used instead.

The mean velocity field is taken as $(0,\,r\Omega,\,\muz)$,
where $\Omega$ is the magnitude of the angular velocity about the galactic center ($\bmOmega=\Omega\hat{z}$ where hat denotes unit vector).
The observable quantities that are parameters of the model are the disk half-thickness $h$, $l$, $u$, $\Omega$,
the shear parameter $q\equiv-\del\ln\Omega/\del\ln r$ ($q=1$ for a flat rotation curve), $\rho$, $\kappa$
and the amplitude $U\f$ of the mean vertical velocity $\muz$. 
For the latter, we adopt the form 
\begin{equation*}
  \muz=U\f z/h.
\end{equation*}
Of these, $\Omega$ and $q$ are typically well-constrained by observation. 
Further, as shown below, the magnetic pitch angle $p=\tan^{-1}(\mbr/\mbp)$ is independent of $\rho$ and $\kappa$.
Moreover, kinematic or steady state solutions can be parameterized in terms of three ratios 
of the four remaining parameters $h$, $l$, $u$, and $U\f$.
For our purposes, it is convenient to consider the parameterization
\begin{equation}
  \label{params}
  H\equiv h/l, \quad \tau\equiv l/u, \quad V\equiv U\f/u.
\end{equation}
Here $H$ is the dimensionless half-thickness of the disk in correlation lengths,
$\tau$ is the eddy turnover time, and $V$ is the dimensionless mean outflow velocity in terms of the rms turbulent velocity.
As we shall see, $\tau$ usually enters the expressions as the inverse Rossby number $\Omega\tau$ 
(also equal to half the Coriolis number),
which is dimensionless, and it is this quantity that we generally work with.
However, it can be interesting to consider $\tau$ separately since $\Omega$ is well-constrained.
An alternate choice is to parameterize the solutions using the dimensionless Reynolds numbers
$R_\alpha\equiv \alpha\f h/\beta$, $R_\Omega\equiv -q\Omega h^2/\beta$, $R_U\equiv U\f h/\beta$, 
where $\alpha\f$ is the amplitude of $\alpha\kin$.

For numerical solutions, we adopt the fairly standard form 
\begin{equation*}
  \alpha\kin=\alpha\f\sin(\pi z/h).
\end{equation*}
We further assume $\tau\corr=\tau$, so that
\begin{equation*}
  \beta=\tfrac{1}{3}lu,
\end{equation*}
and take \citep{Krause+Radler80,Ruzmaikin+88, Brandenburg+Subramanian05a,Brandenburg+13}
\begin{equation}
  \label{Krause}
  \alpha\f=Cl^2\Omega/h,
\end{equation}
where $C$ is a constant close to unity; below we set $C=1$.
It is easy to transform to $R_\alpha$--$R_\Omega$--$R_U$ parameter space using the relations
\begin{equation*}
  \label{param_transform}
  H=\left(\frac{|R_\Omega|}{qR_\alpha}\right)^{1/2},\quad 
  \Omega\tau=\frac{R_\alpha}{3}, \quad 
  V=\tfrac{1}{3}R_U\left(\frac{qR_\alpha}{|R_\Omega|}\right)^{1/2},
\end{equation*}
but below we work in $H$-$\Omega\tau$-$V$ space.
An important quantity is the dynamo number, defined as
\begin{equation}
  \label{Dynamo_number}
  D\equiv R_\alpha R_\Omega=-q(3h\Omega/u)^2=-q(3H\Omega\tau)^2.
\end{equation}
If the $\alpha$ term in the $\del\mbp/\del t$ equation can be neglected (the $\alpha$--$\Omega$ approximation),
solutions in the linear regime are governed by the parameters $D$ and $R_U$ \citep{Ruzmaikin+88}.

It is possible to parameterize equations~\eqref{dynamo} and \eqref{dalpha_mdt} in terms of our chosen parameters. 
To see this, write lengths in terms of $h$ (e.g. $\ztilde\equiv z/h$) 
and magnetic fields in terms of $B\f$ ($\widetilde{\bm{B}}\equiv \meanv{B}/B\f$), 
and define $R_\kappa\equiv\kappa/\beta$ to obtain
\begin{equation*}
  \label{dynamo_param}
  \begin{split}
    \frac{\del\bmBtilde}{\del t}=&\frac{1}{H\tau}\bmDeltilde\cro\bigg[(V\ztilde\zhat+H\Omega\tau\rtilde\phihat)\cro\bmBtilde\\
                                 &+\left(\frac{\Omega\tau}{H}\sin(\pi\ztilde)+\frac{\alp\magn}{u}\right)\bmBtilde
			         -\frac{1}{3H}\bmDeltilde\cro\bmBtilde\bigg],
  \end{split}			        
\end{equation*}
\begin{equation*}
  \begin{split}
    &\frac{\del}{\del t}\left(\frac{\alp\magn}{u}\right)=\\
                              &-\frac{2}{3\tau}\left[\left(\frac{\Omega\tau}{H}\sin(\pi\ztilde)+\frac{\alpha\magn}{u}\right)\Btilde^2
                               -\frac{1}{3H}\bmDeltilde\cro\bmBtilde\cdot\bmBtilde\right]\exp(\ztilde^2)\\
                              &-\frac{1}{H\tau}\bmDeltilde\cdot\left[(V\ztilde\zhat+H\Omega\tau\rtilde\phihat)\frac{\alp\magn}{u}
                     	       -\frac{R_\kappa}{3H}\bmDeltilde\left(\frac{\alp\magn}{u}\right)\right].
  \end{split}			        
\end{equation*}
Thus, if we are interested in averages of $\bmBtilde$ over the disk cross section, 
the solutions are completely determined if $H$, $\tau$, $V$, $\Omega$, $q$, and $R_\kappa$ are specified.
For the reader's convenience, we list in Table~\ref{tab:params} the key parameters of the models.
A typical or `canonical' estimate, e.g. for the solar neighborhood \citep{Ruzmaikin+88,Shukurov04}, 
is given in the rightmost column.

An asymptotic steady state solution for the saturated regime 
can be obtained algebraically by neglecting the $\alpha$ term in the $\phi$--component of Equation~\eqref{dynamo}
(the $\alpha$--$\Omega$ approximation),
replacing $z$-derivatives by divisions by $h$
\citep[with suitable numerical coefficients;][]{Subramanian+Mestel93,Phillips01,Chamandy+14b},
as well as setting time derivatives to zero.
Alternatively, the equations can be solved algebraically in the kinematic (linear) regime by writing $\del\meanv{B}/\del t=\gamma\meanv{B}$,
where $\gamma$ is the exponential growth rate, to obtain $\gamma$ and the kinematic magnetic pitch angle $p\kin$.
Below we summarize the asymptotic solution; details can be found in \citet{Chamandy+14b}.
This solution is useful as a first line of attack and to gain insight into the problem;
however, as explained below, it becomes inaccurate for some regions of parameter space.

\subsection{Asymptotic Solution}
We now summarize the key expressions from the analytical no-$z$ solution of \citet{Chamandy+14b}.
Below, quantities relating to the magnetic field, e.g. $p$, $\mbr$, and $\mbp$, 
refer to the steady state (saturated) values,
whereas the magnetic pitch angle in the kinematic regime is denoted by $p\kin$.
The no-$z$ solution introduces a numerical coefficient $C_U$ of the advective term in Equation~\eqref{dynamo}.
In \citet{Chamandy+14b}, $C_U=1/4$ was used, 
but $C_U=1/2$ gives a much better fit to the numerically determined pitch angles (see Section~\ref{sec:results}),
and so we use $C_U=1/2$ when evaluating expressions.
Growing solutions exist if $D$ is larger in magnitude than the critical dynamo number,
\begin{equation}
  \label{Dcrit}
  D\crit= -\frac{\pi}{32}\left(\pi^2 +4C_UR_U\right)^2=-\frac{\pi}{32}\left(\pi^2 +12C_UH V\right)^2.
\end{equation}
The magnetic pitch angle $p$ in the steady (saturated) state is given by
\begin{equation}
  \label{psat}
  \tan p= \frac{1}{R_\Omega}\left(-\frac{2D\crit}{\pi}\right)^{1/2}
        =-\frac{\pi^2+12C_UH V}{12qH^2\Omega\tau}.
\end{equation}
Note that outflows lead to an increase in $|p|$ because they suppress dynamo action,
causing $|D\crit|$ to be higher.
Note also that we have not made explicit use of the dynamical quenching nonlinearity \eqref{dalpha_mdt},
which may be invoked to obtain an expression for the steady state field strength $\mean{B}$,
which is not needed in the present work.
Thus, solutions for $p$ are not sensitive to the precise form of the dynamo nonlinearity \citep{Chamandy+14b}.

The growth rate of the dynamo in the kinematic regime is given by
\begin{equation}
  \label{gamma}
  \begin{split}
    \gamma&= \sqrt{\frac{2}{\pi}}t\diff^{-1}\left(\sqrt{-D}-\sqrt{-D\crit}\right)\\
          &= \left(\frac{2q}{\pi}\right)^{1/2}\frac{\Omega}{H}
               \left[1 -\left(\frac{\pi}{288q}\right)^{1/2}\left(\frac{\pi^2+12C_UH V}{H\Omega\tau}\right)\right],
  \end{split}
\end{equation}
where $t\diff=h^2/\beta=3H^2\tau$ is the vertical diffusion time.
Note that $D<0$ for $q>0$, that the dynamo is supercritical for $D/D\crit>1$, 
and that larger $|D|$ corresponds to stronger dynamo action.
The pitch angle $p\kin$ in the kinematic regime is given by
\begin{equation}
  \label{pitch_kinematic}
  \tan p\kin=-\left(-\frac{2R_\alpha}{\pi R_\Omega}\right)^{1/2}=-\left(\frac{2}{\pi q}\right)^{1/2}\frac{1}{H}.
\end{equation}
Dividing the first equality of~\eqref{pitch_kinematic} by the first equality of~\eqref{psat}, 
and using the definition of the dynamo number~\eqref{Dynamo_number} gives
\begin{equation}
  \label{Pi}
  \frac{\tan p\kin}{\tan p}=\left(\frac{D}{D\crit}\right)^{1/2},
\end{equation}
so $p$ is similar in both regimes for a mildly critical dynamo.
\begin{table*}
  \begin{center}
    \caption{List of Key Parameters.
             \textbf{Note.} Typical estimates, sometimes referred to as ``canonical" in the text, refer to the solar neighborhood.
            }
    \label{tab:params}
    \begin{tabular}{lccccccc}
      \hline
      Description                                &Symbol     &Units           &Expression                 &Typical Estimate  \\
      \hline                                                                                              
      Equipartition strength at the midplane     &$B\f$      &$\mkG$          &--                         &--                \\
      Magnitude of mean angular velocity         &$\Omega$   &$\kmskpc$       &--                         &$30$              \\
      Radial shear parameter                     &$q$        &--              &$-\del\ln\Omega/\del\ln r$ &$1$               \\
      Mean vertical velocity at the disk surface &$U\f$      &$\kms$          &                           &$1$               \\
      Disk half-thickness                        &$h$        &$\kpc$          &--                         &$0.5$             \\
      Rms velocity of the turbulence             &$u$        &$\kms$          &--                         &$10$              \\
      Dynamo number                              &$D$        &--              &$-q(3\Omega h/u)^2$        &$-20$             \\
      Correlation length of the turbulence       &$l$        &$\kpc$          &--                         &$0.1$             \\
      Eddy turnover time                         &$\tau$     &$\Myr$          &$l/u$                      &$10$              \\
      Turbulent diffusivity of $\meanv{B}$       &$\beta$    &$10^{26}\cmcms$ &$\tfrac{1}{3}lu$           &$1$               \\ 
      Turbulent diffusivity of $\alpha\magn$     &$\kappa$   &$10^{26}\cmcms$ &--                         &$1$               \\
      Dimensionless disk half-thickness          &$H$        &--              &$h/l$                      &$5$               \\
      Inverse Rossby number                      &--         &--              &$\Omega\tau$               &$0.3$             \\
      Dimensionless vertical velocity amplitude  &$V$        &--              &$U\f/u$                    &$0.1$             \\
      Ratio of turbulent diffusivities           &$R_\kappa$ &--              &$\kappa/\beta$             &$1$               \\
      \hline
    \end{tabular}
  \end{center}
\end{table*}

\section{Theoretical constraints}
\label{sec:theor_constraints}
\begin{figure*}                     
  \includegraphics[width=58mm,clip=true,trim=  0 0 10 10]{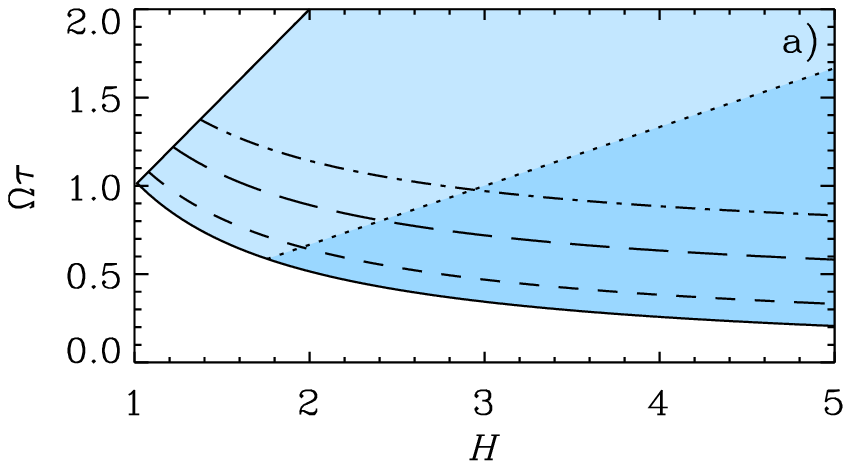}
  \includegraphics[width=58mm,clip=true,trim=  0 0 10 10]{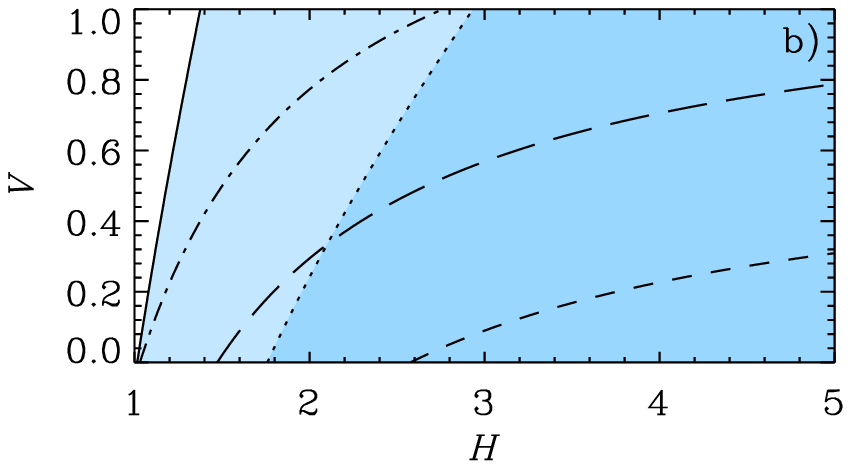}
  \includegraphics[width=58mm,clip=true,trim=  0 0 10 10]{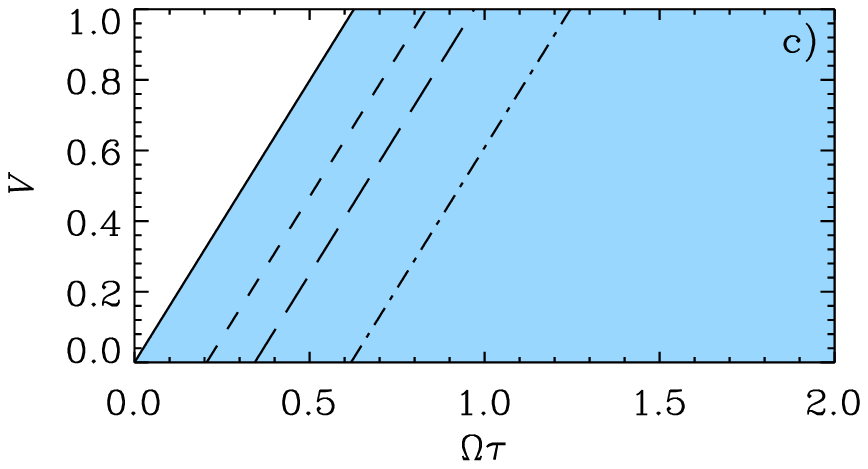}
  \caption{Shaded regions denote the 2-dimensional projections of the 3-dimensional parameter space that is permitted for growing solutions
           under theoretical constraints~\eqref{growing}-\eqref{outflow}
           or \eqref{growing_param}-\eqref{outflow_param}.
           For ease of presentation, $q=1$ is assumed.
           Dotted lines show how the overall upper limits in panels (a) and (b) become modified when $K=1/3$ instead of $1$.
           The allowed parameter space under the more restrictive conditions with $K=1/3$ is shaded darker.
           \textit{Panel~(a)}: Solid lines are from the double inequality~\eqref{growing+realizability+outflow_Phi}. 
           The lower limit~\eqref{growing+realizability_Phi} becomes more restrictive for $V>0$.
           Slices shown are for $V=0.2$ (short dashed), $V=0.6$ (long dashed) and $V=1$ (dashed-dotted).
           \textit{Panel~(b)}: The solid line is from the inequality~\eqref{growing+outflow+realizability_V}.
           The overall upper limit is equivalent to inequality~\eqref{growing+outflow_V} with $\Omega\tau=H$.
           The upper limit becomes more restrictive for other values of $\Omega\tau$:
           $\Omega\tau=0.4$ (short dashed), $\Omega\tau=0.7$ (long dashed), and $\Omega\tau=1$ (dashed-dotted). 
           \textit{Panel~(c)}: the solid line is from the inequality~\eqref{growing+finite_V}.
           The upper limit is in general given by inequality~\eqref{growing+outflow_V},
           and becomes more restrictive for finite values of $H$:
           $H=5$ (short dashed), $H=3$ (long dashed), and $H=5/3$ (dashed-dotted). 
           Colour figures are available in the electronic version.
           \label{fig:theor_constraints}
          }            
\end{figure*}                       

Are there additional constraints on the dynamo equations,
and to what extent do such constraints restrict the available parameter space?
First, as noted above, the dynamo number must have a magnitude that is greater than or equal to the critical value, 
\begin{equation}
  \label{growing}
  |D|\ge|D\crit|;
\end{equation}
otherwise, the magnetic field decays in the linear regime.
By substituting inequality \eqref{growing} into Equation~\eqref{Pi} and making use of Equation~\eqref{pitch_kinematic},
we find
\begin{equation}
  \label{p_growing_Lambda}
  \tan|p|\le\tan|p\kin|=\left(\frac{2}{\pi q}\right)^{1/2}\frac{1}{H}.
\end{equation}
Relation~\eqref{p_growing_Lambda} tells us that $|p|$ in the saturated regime 
cannot exceed its value in the kinematic regime \citep{Elstner05,Chamandy+14b}.
Thus, although outflows help to increase $|p|$, the latter is limited by $|p\kin|$, 
which is independent of $\muz$ in the asymptotic solution.

Second, the energy density of the helical part of the turbulence cannot exceed the total energy density,
\begin{equation}
  \label{realizability}
  \alpha\kin\le Ku,
\end{equation}
where $K$ is a constant of order unity \citep{Moffatt78,Ruzmaikin+88,Brandenburg+Subramanian05a}.
This is sometimes referred to as the realizability condition \citep[but see][]{Ruzmaikin+88}.
Below we retain $K$ in the equations but for numerical examples we adopt $K=1$.
Based on equation \eqref{alpha}, one might argue for the stronger condition with $K=1/3$,
and in some figures we show this case for comparison,
but we choose to be liberal regarding the allowed parameter space given the approximate nature of the model.

Third, we expect the mean vertical velocity to correspond to outflow rather than inflow
\citep{Shukurov+06,Gent+13a,Lagos+13,Chisholm+14},
and so we require
\begin{equation}
  \label{outflow}
  \muz\ge0.
\end{equation}
We are aware that downward turbulent pumping of $\meanv{B}$ 
would lead to a term with the same form as a downward advective term in Equation~\eqref{dynamo},
and may thus have an important effect on the magnetic pitch angle \citep{Gressel10}.
Moreover, some galaxies, especially at high redshift, might accrete, 
but such complications are left for future work.

If conditions \eqref{growing}, 
\eqref{realizability} and \eqref{outflow} are expressed in terms of the dimensionless parameters of interest,
we have
\begin{equation}
  \label{growing_param}
   q(3H\Omega\tau)^2\ge \frac{\pi}{32}\left(\pi^2+12C_UH V\right)^2, 
\end{equation}
\begin{equation}
  \label{realizability_param}
  KH\ge\Omega\tau,
\end{equation}
\begin{equation}
  \label{outflow_param}
  V\ge0.
\end{equation}
Below, we also implicitly make use of the facts that $H>0$ and $\tau>0$ and assume $q>0$.

It is useful to rewrite these theoretical constraints as constraints on individual parameters. 
Solving for $H$ from relation~\eqref{growing_param}, 
we obtain a lower limit for $H$ valid for all $\Omega\tau$ and $V$,
\begin{equation}
  \label{growing_Lambda}
  H\ge\left(\frac{\pi^5}{288q}\right)^{1/2}\left[\Omega\tau-\left(\frac{\pi}{2q}\right)^{1/2}C_UV\right]^{-1}.
\end{equation}
On the other hand, obtaining a lower limit for $\Omega\tau$ from inequality~\eqref{growing_param}, 
and combining this with the upper limit~\eqref{realizability_param},
we obtain
\begin{equation}
  \label{growing+realizability_Phi}
  \left(\frac{\pi^5}{288q}\right)^{1/2}\left(\frac{1}{H}+\frac{12C_UV}{\pi^2}\right)\le\Omega\tau\le KH.
\end{equation}
Given that the left hand side must be less than or equal to the right hand side, we deduce a lower limit on $H$ in terms of $V$ only:
\begin{equation}
  \label{growing+realizability_Lambda}
  H\ge\left(\frac{\pi}{8q}\right)^{1/2}\frac{1}{K}\left\{C_UV+\left[(C_UV)^2+\frac{K}{3}(2\pi^3q)^{1/2}\right]^{1/2}\right\}.
\end{equation}
To obtain an overall lower limit on the inverse Rossby number $\Omega\tau$, 
we can incorporate condition~\eqref{outflow_param} into expression~\eqref{growing+realizability_Phi},
so that for all $V\ge0$, we have
\begin{equation}
  \label{growing+realizability+outflow_Phi}
  \left(\frac{\pi^5}{288q}\right)^{1/2}\frac{1}{H}\le\Omega\tau\le KH.
\end{equation}
From double inequality~\eqref{growing+realizability+outflow_Phi} we also obtain an overall limit on $H$,
\begin{equation}
  \label{growing+outflow+realizability_Lambda}
  H\ge\left(\frac{\pi^5}{288q}\right)^{1/4}\frac{1}{K^{1/2}},
\end{equation}
where the right hand side evaluates to $1.0$ for a flat rotation curve ($q=1$) with $K=1$.
This result is consistent with the expectation that the turbulent scale $l$ should not exceed $\sim h$ for 3-dimensional turbulence.
Combining constraint~\eqref{growing+outflow+realizability_Lambda} with inequality~\eqref{p_growing_Lambda}
leads directly to an upper limit to the pitch angle, $\tan|p|\le[(9/q)(2/\pi)^7]^{1/4}K^{1/2}\simeq0.79$ for $q=1$ and $K=1$, 
or $\simeq38^\circ$.
For $q=0.6$, this upper limit rises to $\simeq42^\circ$.
Of the 12 galaxies listed in the \citet{Vaneck+15} data,
the only galaxy observed to have $|p|$ larger than the above limit is M33, which has $p\sim-(36$--$60)^\circ$ 
(where the range includes radial variations as well as 1$\sigma$ errors) \citep{Tabatabaei+08,Vaneck+15}.
This galaxy also has rather small angular velocity $\Omega\sim25-41\kmskpc$ and radial shear $q\sim0.6-0.8$ \citep{Sofue+99,Vaneck+15}.

Double inequality~\eqref{growing+realizability+outflow_Phi} governs the allowed $H$--$\Omega\tau$ parameter space,
which is a 2-dimensional projection of the 3-dimensional $H$--$\Omega\tau$--$V$ parameter space.
This parameter space is represented by the shaded region in Figure~\ref{fig:theor_constraints}a, where $q=1$ and $K=1$ have been adopted.
A dotted line is also included to show how the overall upper limit ($V=0$) shifts if $K=1/3$ 
is adopted rather than the less conservative $K=1$.
The accessible parameter space under the more restrictive value $K=1/3$ is denoted by darker shading.
If $V$ is specified (and is other than $0$), then relation~\eqref{growing+realizability_Phi} must be used, 
and the lower limit gets adjusted upwards, 
as shown by the short dashed ($V=0.2$), 
long dashed ($V=0.6$), and dashed-dotted ($V=1$) curves.
These represent slices of the 3-dimensional parameter space.

We now derive constraints on the velocity ratio $V$. 
We obtain an upper limit by rearranging expression~\eqref{growing_param},
and then combine this with the lower limit~\eqref{outflow_param} to obtain
\begin{equation}
  \label{growing+outflow_V}
  0\le V\le\left(\frac{2q}{\pi}\right)^{1/2}\frac{\Omega\tau}{C_U}-\frac{\pi^2}{12C_UH}.
\end{equation}
Making use of relation~\eqref{realizability_param},
the upper limit can be generalized to apply for all $\tau$:
\begin{equation}
  \label{growing+outflow+realizability_V}
  0\le V\le\left(\frac{2q}{\pi}\right)^{1/2}\frac{KH}{C_U}-\frac{\pi^2}{12C_UH}.
\end{equation}
This allowed $V$--$H$ parameter space (for $q=1$ and $K=1$) is shown shaded in Figure~\ref{fig:theor_constraints}b.
Expression~\eqref{growing+outflow_V} can be used to draw slices of this parameter space,
that is to adjust the upper limit downwards for specific values of $\Omega\tau\le H$.
These are represented by the short dashed ($\Omega\tau=0.4$), long dashed ($\Omega\tau=0.7$), and dashed-dotted ($\Omega\tau=1$) curves.

Finally, we consider the allowed region of $V$--$\Omega\tau$ space.
For this purpose we require an upper limit on $H$, which,
to be liberal, is simply taken to be $\infty$ (in practice we are more interested in small $H$ anyway).
Thus, $-\pi^2/(12C_UH)<0$ in the right hand side of expression~\eqref{growing+outflow_V},
so that for all $H$ we have,
\begin{equation}
  \label{growing+finite_V}
  0\le V<\left(\frac{2q}{\pi}\right)^{1/2}\frac{\Omega\tau}{C_U}.
\end{equation}
Thus any upper limit on $\tau$ translates to an upper limit on $V$.
In Figure~\ref{fig:theor_constraints}c, the allowed region of $V$--$\Omega\tau$ parameter space has been shaded.
The solid line corresponds to an upper limit on $V$ for $H\rightarrow\infty$,
while revised upper limits for a sample of finite values of $H$ are also plotted:
$H=5$ (short dashed), $H=3$ (long dashed), and $H=5/3$ (dashed-dotted).

The physical constraints~\eqref{growing}, \eqref{realizability}, and \eqref{outflow} could be recast in terms of $p$,
using Equation~\eqref{psat}, and then observations of $\Omega$, $q$, and $p$ used to constrain the $H$--$\tau$--$V$
parameter space for each data point.
However, we leave a more detailed comparison of theoretical solutions and observational data for a future work.
We now turn to the main results dealing with theoretical predictions for the magnetic pitch angle.

\section{Predictions for the magnetic pitch angle}
\label{sec:results}
\begin{figure}
  \includegraphics[height=28.79mm,clip=true,trim=  0 20 3   0]{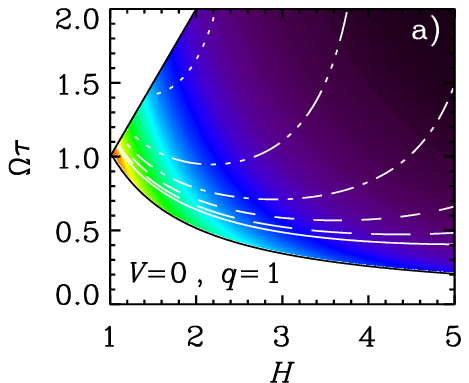}	
  \includegraphics[height=28.79mm,clip=true,trim=  5 20 5   0]{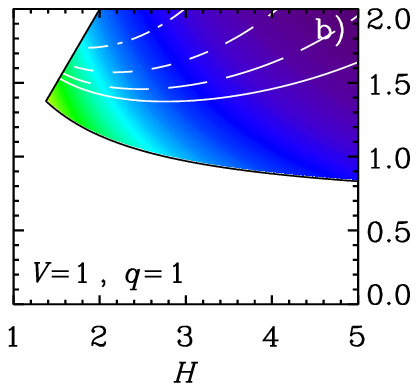}\\	
  \includegraphics[height=35.00mm,clip=true,trim=  0  0 3   0]{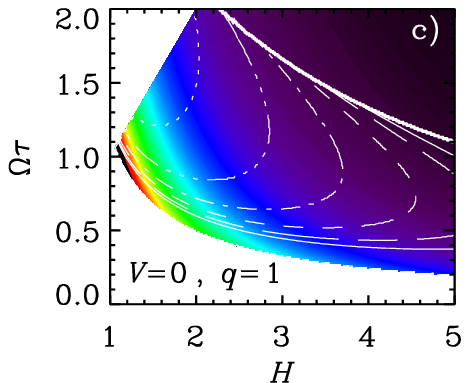}	
  \includegraphics[height=35.00mm,clip=true,trim=  5  0 5   0]{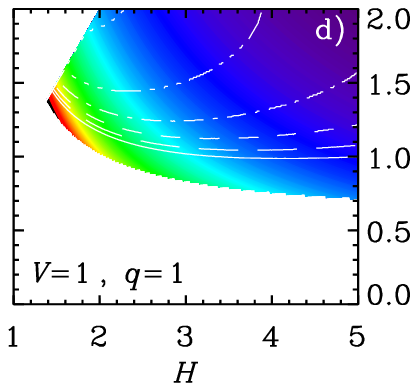}\\
  \includegraphics[width= 80.00mm,clip=true,trim=-32 12 0 -10]{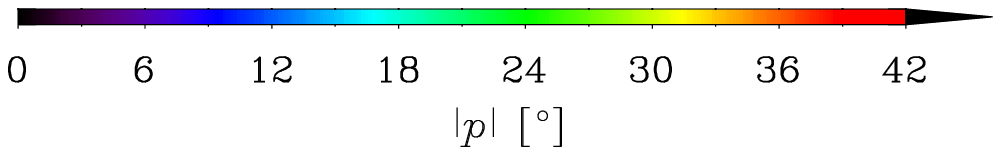}
  \caption{Magnetic pitch angle magnitude, shown as a function of the parameters $H$ and $\Omega\tau$,
           for asymptotic solution (top row) and numerical solutions (bottom row),
           for the case $q=1$, and for $V=0$ (left column), and $V=1$ (right column).
           Parameter space for which $|p|>42^\circ$ [left side of panels (c) and (d)] is colored black.
           Contours denote $1000$-folding times, in galactic rotation periods $T\equiv2\pi/\Omega$,
           in increments of $2T$ with dotted curves corresponding to $4T$ and solid curves corresponding to $14T$.
           The thick solid line separates the regions for which the kinematic solution is steady (below this line)
           or oscillatory (above this line).
           Uncolored regions designate disallowed parameter space.
           \label{fig:p_V}
          }            
\end{figure}                       
\begin{figure}
  \includegraphics[height=28.79mm,clip=true,trim=  0 20 3   0]{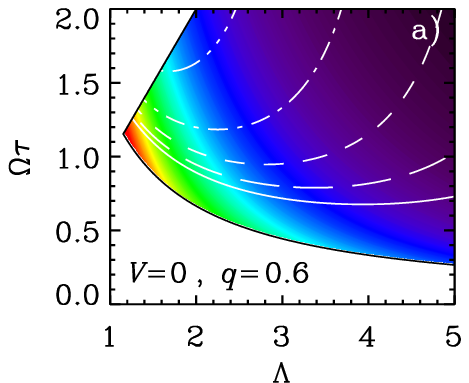}	
  \includegraphics[height=28.79mm,clip=true,trim=  5 20 5   0]{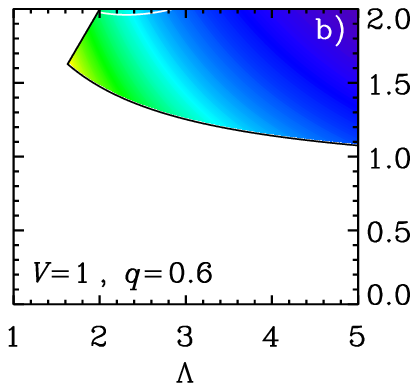}\\	
  \includegraphics[height=35.00mm,clip=true,trim=  0  0 3   0]{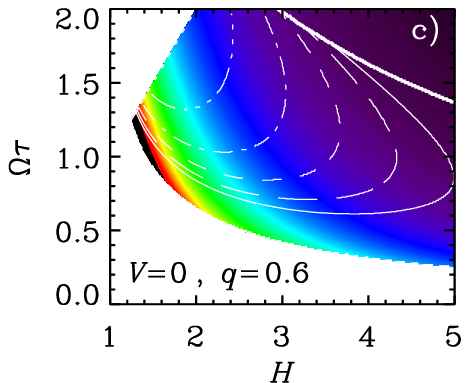}
  \includegraphics[height=35.00mm,clip=true,trim=  5  0 5   0]{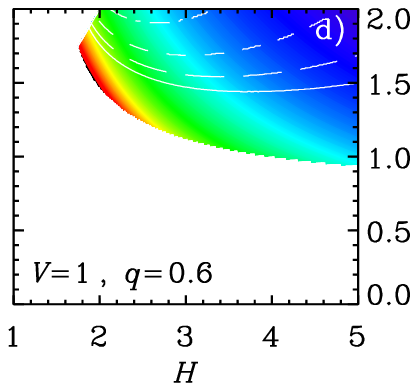}\\
  \includegraphics[width= 80.00mm,clip=true,trim=-32 12 0 -10]{p_colorbar.eps}
  \caption{As Figure~\ref{fig:p_V} but for $q=0.6$.
           \label{fig:p_V_q06}
          }            
\end{figure}                       
We employ numerical as well as analytic methods and compare the values of $p$ from the resulting solutions.
Closed-form asymptotic analytical solutions can be useful, but it is important to be aware of their limitations,
and numerical solutions provide an important check.
Moreover, as mentioned above, the parameter $C_U$ in the asymptotic solution was calibrated using the numerical solution.
As discussed below, for certain regions of parameter space the two types of solution differ
because certain approximations that go into the analytic method become inaccurate,
but generally, the two types of solution are mutually consistent so far as predictions of the pitch angle are concerned.
One minor difference that deserves mentioning is that for numerical solutions, constraint~\eqref{growing} is imposed only implicitly,
as the field will decay if $|D|<|D\crit|$, while for analytical solutions, it is imposed explicitly.
In practice, $D\crit$ for the numerical solutions will differ slightly from the analytic estimate~\eqref{Dcrit} 
and must be calculated by iterating until $\gamma\approx0$.
On the other hand, constraints~\eqref{realizability}/\eqref{realizability_param} and \eqref{outflow}/\eqref{outflow_param} 
are applied explicitly in the numerical model as in the analytic model,
that is numerical solutions are not calculated if these constraints are not satisfied.

Pitch angles are obtained from numerical solutions using the estimate $\langle p\mean{B}^2\rangle/\langle\mean{B}^2\rangle$,
where angular brackets represent the mean value of grid points between $-h$ and $h$.
We weight the pitch angle by $\mean{B}^2$ because the intensity of polarized synchrotron emission $P$ also scales as $\mean{B}^2$.
We note that this represents a conservative estimate, as $|p|$ at the midplane is typically larger than this average value.
In fact, to be precise, $P\propto\int_L n\cray \mean{B}_\perp^2 ds$ 
where $n\cray$ is the number density of relativistic (cosmic ray) electrons, 
$L$ is the path length from source to observer,
and $\mean{B}_\perp$ is the component of $\meanv{B}$ in the plane of the sky
\citep[e.g.][]{Shukurov04}.\footnote{This equation ignores the contribution of the anisotropic random component of the magnetic field, 
but this does not have any bearing on the present discussion.}
If we make the approximation \citep{Beck+96} that $n\cray\propto B^2$, 
where $B$ is the strength of the total (mean$+$random) magnetic field,
and assume $B\propto B\eq$ \citep[][see also the Johns Hopkins Turbulence Database, 
\citealt{Li+08}]{Mckee+Zweibel95,Fletcher+Shukurov01,Basu+Roy13,Gent+13b},
then the pitch angle should be weighted by the extra factor $B\eq^2$.
We have redone the calculations including this extra factor but find that $|p|$ increases by up to only $10\%$,
or at most $3^\circ$ for the parameter space considered.
In the discussion below, we adopt the more conservative flat ($z$--independent) profile for $n\cray$.

Figure~\ref{fig:p_V} shows color plots of $|p|$ as a function of $H$ and inverse Rossby number $\Omega\tau$,  
from analytical (Equation~\eqref{psat}, top panels (a) and (b)) and numerical solutions (bottom panels (c) and (d)),
for the case $q=1$.
Numerical solutions were calculated at steps of $0.01$ in $H$ and in $\Omega\tau$,
with each point corresponding to a separate dynamo run with $21$--$41$ grid points (excluding ghost zones), 
as needed. 
A subset of solutions were checked for consistency against higher resolution runs.
As expected, numerical solutions for $p$ in the steady state, as well as for $p\kin$ in the linear regime, 
are insensitive to the value of $R_\kappa$, which is generally set to $1$.
The slice $V=0$ is shown in panel~(a) or (c) while $V=1$ is shown in panel~(b) or (d).
Numerical solutions are generally well-approximated by asymptotic solutions,
but it can be seen that asymptotic solutions tend to underestimate $|p|$ and overestimate $|D\crit|$.
Further, the level of agreement between the two types of solution decreases as $V$ increases.
From the figure it is evident that theoretical pitch angles of magnitude $\gtrsim20^\circ$ can be attained, 
but require values of $H$ smaller than canonical, i.e. smaller than $\sim4$--$5$.
Moreover, as seen by comparing Figures~\ref{fig:p_V}a and \ref{fig:p_V}b, 
$|p|$ is larger for given values of $H$ and $\tau$ when $V$ is larger.
However, clearly $\Omega\tau$ must be large enough to ensure that the dynamo remains supercritical,
a constraint that is more easily satisfied for smaller $V$.
We emphasize that since $\Omega$ is generally well constrained by observations, 
it makes sense to discuss the allowed range of $\tau$.
For example, if $\Omega\tau=1$, then for $\Omega=40\kmskpc$, $\tau=0.024\Gyr$,
which is somewhat larger than $\tau=0.01\Gyr$,
obtained from the standard estimates $l=0.1\kpc$ and $u=10\kms$.

We also overplot contours for the $1000$-folding time $t_{1000}$
(time for the magnetic field to amplify by the factor $10^3$ in the kinematic regime)
in units of the galactic rotation period $T=2\pi/\Omega$.
Dotted curves correspond to $t_{1000}=4T$ and contours are spaced by $2T$.
It can be seen that growth times are over-estimated by analytical solutions,
especially for $V>0$.
The \textit{thick} solid line in panel (c) is the dividing line between two different regimes.
Above this line,
kinematic solutions are oscillatory,
but become steady upon saturation.
This oscillatory behavior is not dependent on the $\alpha^2$ effect nor is it sensitive to the form of $\alpha\kin$;
the fact that it was not seen in solutions of \citet{Brandenburg98,Brandenburg+Subramanian05a} 
is probably due to the symmetry conditions imposed at the midplane in their model,
which are different from the ones which arise naturally in our solutions.
In any case, this regime is not particularly relevant for the present work,
but its investigation in a future study would be interesting.

The case $q=0.6$ is illustrated in Figure~\ref{fig:p_V_q06}.
Pitch angles are somewhat larger in magnitude compared to $q=1$.
This is expected because the magnitude of the $\Omega$-effect is reduced compared to that of the $\alpha$-effect.
However, numerical solutions have noticeably larger $|p|$ than analytical solutions for $H\sim1-2$.
As discussed below, this is mainly due to the $\alpha^2$ effect,
which is neglected in the analytic model.

\begin{figure}
  \includegraphics[width=85mm,clip=true,trim=  10 30 0 10]{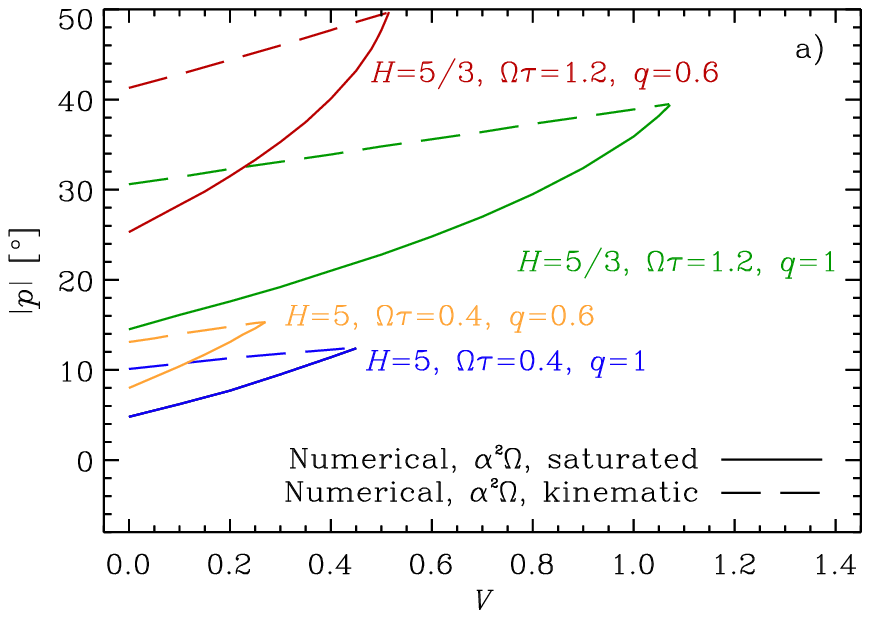}\\
  \includegraphics[width=85mm,clip=true,trim=  10  0 0 10]{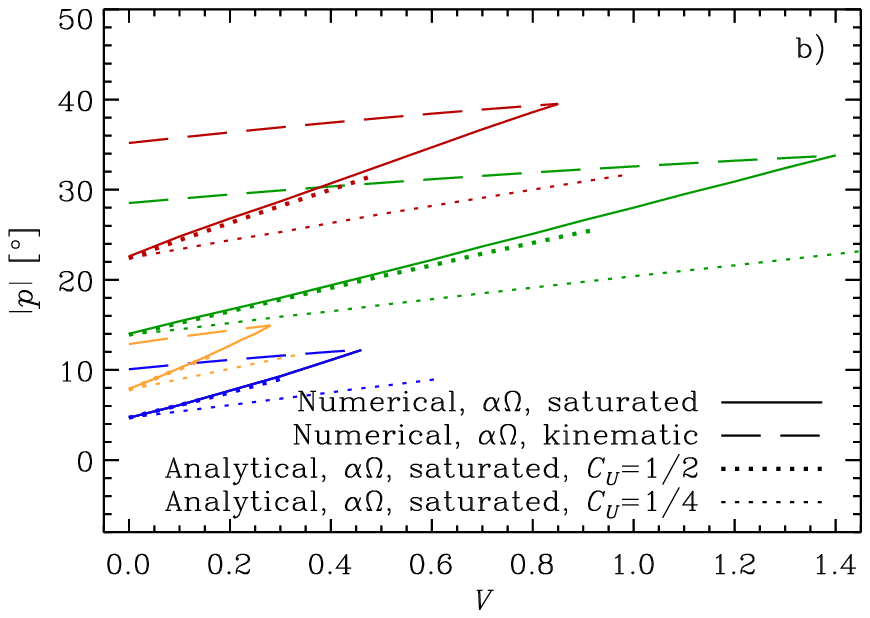}
  \caption{\textit{Panel (a)}: Full $\alpha^2$--$\Omega$ numerical saturated solution (solid), 
           as well as full numerical kinematic solution (dashed),
           for four different sets of parameters,
           distinguished by color as indicated on the plot.
           All curves have $\Omega=40\kmskpc$ and $R_\kappa=1$ (though solutions are insensitive to $R_\kappa$). 
           Lines are drawn from $V=0$ up until the point where the dynamo becomes critical.
           \textit{Panel (b)}: Numerical saturated solution with $\alpha$--$\Omega$ approximation (solid),
           along with the corresponding kinematic solution (dashed).
           Analytical steady state solutions are shown for $C_U=1/2$ (thick dotted) and $C_U=1/4$ (thin dotted).
           For both values of $C_U$, $|p|$ rises to the same value $|p\kin|$ before the solution goes subcritical.
           \label{fig:p_vs_V}
          }            
\end{figure}

Figure~\ref{fig:p_vs_V} compares analytical and numerical solutions for four sets of parameters,
with each set represented with a different color.
In both panels, curves cut off for the value of $V$ above which the dynamo is subcritical (where it can be seen that $p=p\kin$).
All curves assume $\Omega=40\kmskpc$ (close to the median value in the \citealt{Vaneck+15} data set).
In panel (a), solid curves show full steady-state numerical solutions,
while dashed curves show the corresponding kinematic solutions.
Occupying the lower portion of the panel in blue, 
we have curves for `canonical' values $H=5$, $\tau=0.01\skpckm\simeq0.01\Gyr$ and $q=1$.
Moving up to the set of orange curves, we have $H=5$, $\tau=0.01\skpckm$ and $q=0.6$.
The green set of curves illustrates $H=5/3$, $\tau=0.03\skpckm\simeq0.03\Gyr$, $q=1$.
Finally, the upper red set of curves shows $H=5/3$, $\tau=0.03\skpckm$, $q=0.6$.
The simplest way to go from the blue to green or orange to red curves is to take $l$ to be larger by a factor of $3$.
Figure \ref{fig:p_vs_V}b shows solutions for which the $\alpha$ term 
was excluded from the $\phi$-component of Equation~\eqref{dynamo} (the $\alpha$--$\Omega$ approximation),
Solid and dashed curves again represent numerical saturated and kinematic solutions, respectively.
We also show analytical solutions with $C_U=1/2$ (the value adopted in calculations, thick dotted), 
and $C_U=1/4$ (thin dotted).
It is apparent that using $C_U=1/2$ leads to a much more accurate estimate of $p$.

Clearly the $\alpha$--$\Omega$ approximation is justified for canonical parameters (blue curves),
and the analytical solution (with $C_U=1/2$) reproduces well the numerical solution,
except that it underestimates the maximum $V$ for which the dynamo remains supercritical
(it overestimates $|D\crit|$ and underestimates $\gamma$).
Going now to the green curves, which have smaller $H$ than the blue curves,
we see that $|p|$ is larger, as expected. 
The asymptotic solution approximates quite well the numerical solution that does not include the $\alpha^2$ effect,
but clearly the $\alpha$--$\Omega$ approximation is inaccurate for large $V$.
Including it leads to much larger $|p|$.
This arises because this term leads to a suppression of $\mbp$ near the midplane.\footnote{In 
principle this effect could be accommodated in the analytic model by replacing $q$ with $q-g$,
where $g$ is a positive function of some of the parameters.
However, the nonlinear dependence of $p$ on $V$ complicates such an approach.}
Finally, moving to the upper red curves with $q=0.6$,
it is evident that $|p|$ is larger than for $q=1$, as expected.
Moreover, the $\alpha^2$ effect plays an even bigger role since the ratio
of the shear term to the $\alpha$ term in the equation for $\del\mbp/\del t$ is proportional to $q$.
We note that if the dynamo is near-critical, the growth rate is small, 
and, depending on the seed field assumed, the magnetic field may not have time to saturate for a given integration time (say $10\Gyr$).
However, this does not mean that large magnitude pitch angles cannot be achieved in such cases,
because $|p|$ approaches monotonically or almost monotonically the saturated value from its value in the kinematic regime $|p\kin|$,
which is, in any case, the upper limit of $|p|$.

\begin{figure}
  \includegraphics[width=85mm,clip=true,trim=  0 30 0 0]{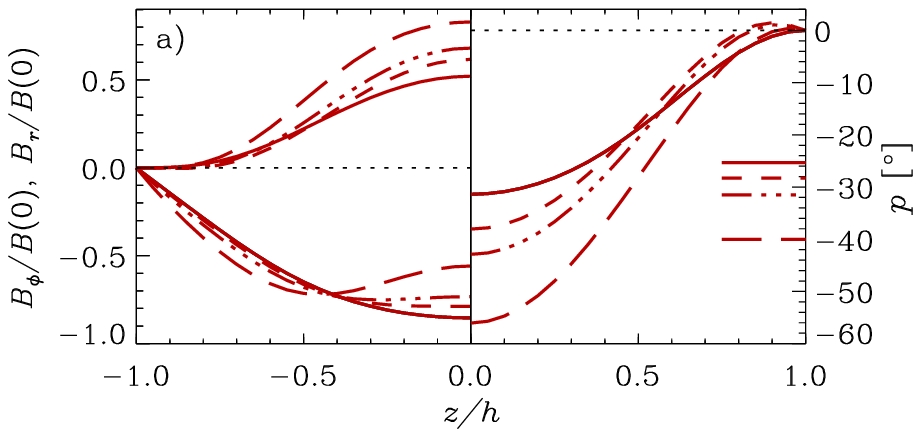}\\
  \includegraphics[width=85mm,clip=true,trim=  0 30 0 0]{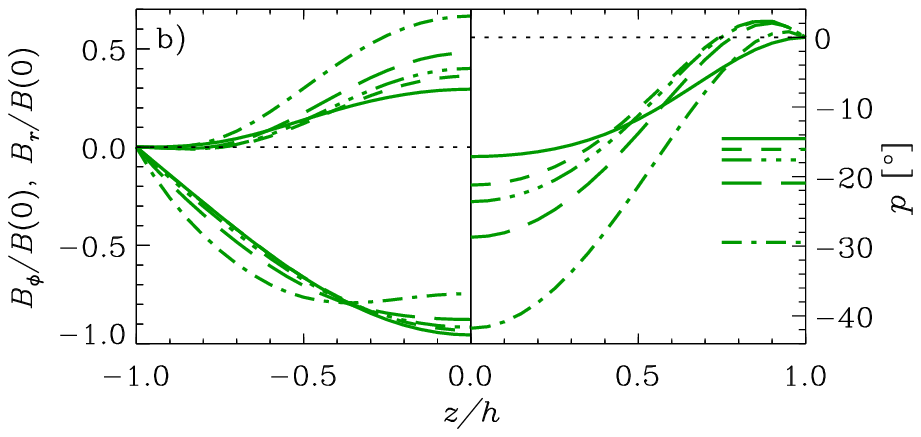}\\
  \includegraphics[width=85mm,clip=true,trim=  0 30 0 0]{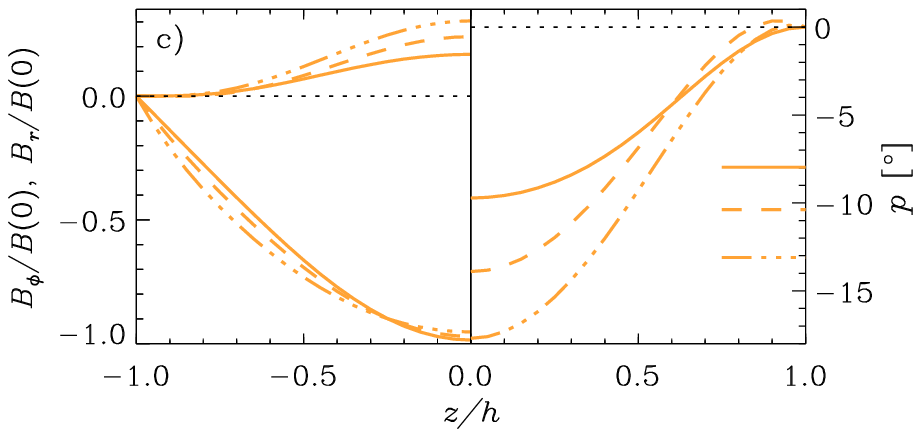}\\
  \includegraphics[width=85mm,clip=true,trim=  0 0  0 0]{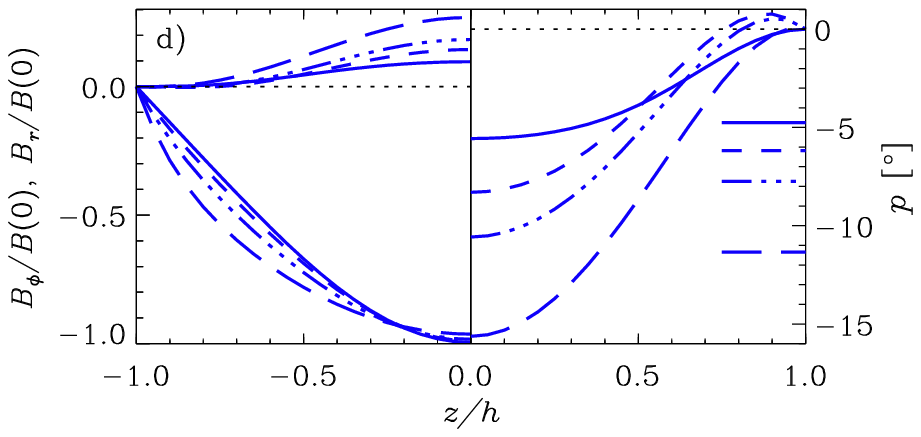}
  \caption{Magnetic field profiles (left) and magnetic pitch angles $p$ (right) for each of the parameter sets
           depicted in Figure~\ref{fig:p_vs_V}, with corresponding color.
           \textit{Panel~(a)}: $H=5/3$, $\tau=0.03\skpckm$, $q=0.6$;
           \textit{Panel~(b)}: $H=5/3$, $\tau=0.03\skpckm$, $q=1$;
           \textit{Panel~(c)}: $H=5$, $\tau=0.01\skpckm$, $q=0.6$;
           \textit{Panel~(d)}: $H=5$, $\tau=0.01\skpckm$, $q=1$.
           Linestyles represent different outflow speeds: 
           $V=0$ (solid), $V=0.1$ (short dashed), $V=0.2$ (dash-triple-dotted),
           $V=0.4$ (long dashed), and $V=0.8$ (dashed-dotted).
           Horizontal lines show the $\mean{B}^2$-weighted spatial mean of $p$.
           \label{fig:B_vs_z}
          }            
\end{figure}

The spatial profile of the magnetic field for each of the parameter sets of Figure~\ref{fig:p_vs_V} is plotted in Figure~\ref{fig:B_vs_z},
using the same color for each parameter set as in Figure~\ref{fig:p_vs_V}.
Curves for the steady state values of $\mbr$ and $\mbp$ are plotted on the left,
while $p$ is plotted on the right,
for $V=0$ (solid), $V=0.1$ (short dashed), $V=0.2$ (dash-triple-dotted), $V=0.4$ (long dashed), and $V=0.8$ (dashed-dotted),
for cases for which such an outflow does not lead to a subcritical dynamo.
All quantities plotted are even about the midplane.
In our solutions, $\mbp$ tends to mainly negative values, while $\mbr$ tends to mainly positive values,
but the equations are invariant under a sign reversal of $\meanv{B}$,
and the signs are determined by the arbitrary seed field chosen.
Both $\mbp$ and $\mbr$ are normalized with respect to the midplane mean field strength $\mean{B}(0)=[\mbr^2(0)+\mbp^2(0)]^{1/2}$.
The average pitch angle $\langle p\mean{B}^2\rangle/\langle\mean{B}^2\rangle$ is also shown as a horizontal line of the appropriate linestyle.
As mentioned above, the largest values of $|p|$ arise for small $H$ and large $\Omega\tau$ and $V$,
when there is a significant reduction of $\mbp$ near the midplane,
a consequence of the $\alpha^2$ effect.

We see then that large magnitude pitch angles can indeed be achieved for small values of $H$
and large values of $\Omega\tau$ and $V$ relative to canonical.
This cannot be fully appreciated by using the asymptotic solution because it does not account for the $\alpha^2$ effect,
and so underestimates $|p|$, especially for the parameter values that are favourable for large pitch angles.
Moreover, past works \citep{Chamandy+14b,Vaneck+15} 
tend to underestimate $|p|$ for non-zero outflows because they use $C_U=1/4$, whereas here we use $C_U=1/2$, 
which is a much better calibration to numerical solutions for making estimates of $p$
(though perhaps worse for making estimates of $\gamma$).
Perhaps the simplest way to achieve large $|p|$ is to make $l$ and $U\f$ larger than standard estimates.
Is such a prescription plausible in light of independent evidence?

\section{What parameter values are realistic?}
\label{sec:discussion} 
The magnetic field is believed to roughly trace the warm ionized gas component \citep{Beck+96}.
Outflow speeds of the hot gas tend to be as large as a few$\times10^2\kms$.
However, $\muz$ represents an average over all interstellar medium (ISM) phases inhabited by the mean magnetic field,
as well as over the disk surface, which includes regions in between hot superbubbles (chimneys),
leading to standard estimates $\muz\sim0.2$--$2\kms$ \citep{Shukurov+06,Vaneck+15,Rodrigues+15}.
Such estimates involve equating the mass flux for the averaged ISM to that of the outflowing hot gas.
However, simulations and observations seem to indicate that warm gas can become entrained in the hot outflows,
leading to velocities of the warm gas of several tens of $\kms$ or more \citep[e.g.][]{Gent+13a,Chisholm+14}.
Therefore, values of $\muz$ comparable with $u\sim10$--$20\kms$ could possibly be more realistic than past estimates,
at least for some galaxies.
It is worth emphasizing, however, 
that an extra term of the form $\vec{\gamma}\cro\meanv{B}$ in Equation~\eqref{emf} for $\Emf$,
with $\gamma_z<0$, could become important when the gaseous halo is included \citep{Gressel10}.
The resulting turbulent pumping of the magnetic field towards the midplane 
would lead to an effective value of $\muz$ in Equation~\eqref{dynamo} that is lower than the actual mean vertical velocity,
and could, in principle, even be negative.
This pumping term would not, however, play any role in Equation~\eqref{dalpha_mdt}, 
so the advective helicity flux would not be directly affected.

The correlation length of turbulence $l$ has been relatively challenging to determine observationally.
Many studies have used Faraday rotation measurements to probe the scale of magnetic field fluctuations.
However, these scales may be different from the scale $l$ \citep{Brandenburg+Subramanian05a}. 
More relevant to the present study, perhaps, are studies that set out to determine the velocity correlation length directly.
For the Milky Way, \citet{Chepurnov+10} find a scale of $0.140\pm0.080\kpc$ for the warm HI.
Observations of nearby dwarf galaxies yield values for $l$ of a few $\!\kpc$ \citep{Elmegreen+01,Stanimirovic+Lazarian01,Dib+Burkert05}.
\citet{Dutta+13} obtain HI power spectra for a sample of 18 spiral galaxies
and find power law fits extending up to a scale comparable to the radial scale length of the optical disk 
(several $\!\kpc$),
with power laws flatter than those found at smaller scales.
They attribute this flattening to a transition from 3-dimensional to 2-dimensional turbulence for $l\sim h$, 
where $h$ is the disk scale height.

Many different drivers of turbulence in the ISM
have been proposed in the literature \citep[][and references therein]{Falceta-goncalves+14},
and more than one such driver may, in fact, be important.
Hydrodynamic simulations of the local ISM with supernova-driven turbulence by \citet{Avillez+Breitschwerdt07} predict a value $l\sim0.075\kpc$,
whereas \citet{Gent+13a} predicts $l\sim0.1\kpc$ at the midplane, rising to $\sim0.3\kpc$ at $|z|=1\kpc$.
These determinations of $l$ were for the averaged ISM in the solar neighborhood, rather than for any particular phase.
Simple analytic estimates based on the size attained by a supernova remnant by the time it mixes with the ambient ISM 
give $l\sim0.1$-$0.2\kpc$ for the warm gas \citep[e.g.][]{Lacki13}. 
However, if the turbulence is driven by expanding supernova-driven superbubbles, 
then the turbulent scale may be comparable to the largest sizes attained by such bubbles before mixing with the ambient ISM.
Such bubbles may become as large as a few $\!\kpc$ in some galaxies \citep[e.g.][]{Boomsma+08}, 
exceeding the scale height of the thin component of the disk,
but still comparable to that of the diffuse ionized medium \citep{Gaensler+08}.
The bubble cross-sectional area at the mixing stage is expected to increase with distance $z$ from the midplane,
as well as with the galactocentric radius $r$ \citep{Ferriere98}.
It is not clear whether the box size of the ISM simulations mentioned would have been large enough to capture turbulence on such large scales.
In a recent work, \citet{Moraghan+15} find in their simulations that
the spherical outflows that drive the turbulence transfer power to scales larger than the initial injection scale as they expand.
Based on all of this evidence, it is not possible to exclude any value for $H\equiv h/l$ in the range $1$-$10$.

Note that raising $l$ has no effect on the dynamo number $D$ (Equation~\eqref{Dynamo_number}), 
though it does have the effect of decreasing the magnitude of the critical dynamo number $D\crit$ (Equation~\eqref{Dcrit}).
Thus, the ratio $D/D\crit$, which governs dynamo growth via Equation~\eqref{gamma},
increases from its canonical value, if other parameters are kept constant.
In fact, it can be seen from Equation~\eqref{gamma}
that the exponential growth rate of the mean magnetic field in the kinematic regime $\gamma$ increases for two reasons.
First, $|D\crit|$ is reduced, and second, the turbulent diffusion time $t\diff\propto1/l$ is reduced.
Thus, a larger $l$ implies a more supercritical dynamo which is able to grow the mean magnetic field more rapidly.
This is also evident in Figures~\ref{fig:p_V} and \ref{fig:p_V_q06}, 
where smaller $H$ and larger $\tau$ correspond to smaller exponential growth times in the kinematic regime.
On the other hand, a model with a larger $l$ can accommodate a larger, perhaps more realistic value of $V$,
for a given growth rate.

The tentative conclusion is that the parameter values required for larger magnitude pitch angles 
are not inconsistent with observational and theoretical constraints.
However, these parameter values (most notably $H\sim1-3$) sit in a small region of the allowed parameter space,
near a `corner' (left-hand sides of the panels of Figures~\ref{fig:p_V} and \ref{fig:p_V_q06}) 
that is formed by rather robust theoretical constraints (`slanted walls'), 
namely the requirement for dynamo growth and the realizability condition.
Therefore, although $|p|\gtrsim20^\circ$, and even $|p|\gtrsim40^\circ$ as in the galaxy M33, is in principle attainable,
the parameter values that allow large pitch angle are at the extreme range of the allowed parameter space.
For the largest possible values of $|p|$, the dynamo must be close to critical
\textit{and} the turbulent velocity field must be nearly maximally helical
(even then if $K=1/3$ instead of $K=1$, $|p|\gtrsim40^\circ$ would no longer be attainable). 
Of course, it is possible that physical mechanisms exist to push parameters into this region of parameter space.
It is also possible that the theory is still missing some important physical ingredients 
that would lead to new effects that would in turn tend to increase the value of $|p|$.
Where should we look?

\section{Possible extensions of the model}
\label{sec:extensions}
There are many possibilities for extending our basic model.
The most obvious is to include the radial dependence.
This would introduce new parameters (radial scale lengths) that would complicate the model.
Past studies that included the radial coordinate 
have shown $p$ to be of roughly the same magnitude as predicted by local solutions \citep{Chamandy+13a}.
To verify this, we have carried out a preliminary study using a mean-field simulation in $r$--`no-$z$', 
with a Brandt rotation curve \citep[see][for details of the model]{Chamandy+13a}.
Outside a small central region (where the thin disk approximation may not be valid anyway)
we generally find that $p(r)$ in the saturated regime is very close to the local analytical prediction,
until $r\gtrsim r\crit$, where $r\crit$ is defined by $D(r\crit)/D\crit(r\crit)\equiv1$.
For values of $r$ up to $\sim1\kpc$ larger than $r\crit$, the magnetic field in the saturated state can still be significant,
and interestingly, we find that here the pitch angle saturates near its kinematic value.
While such issues are beyond the scope of this paper, they deserve further study.
Specifically, detailed comparisons of solutions obtained for local models in $z$ 
and global models in $r$--$z$ with and without the thin disk approximation would be useful.
It would also be interesting to compare theoretical predictions for the variation of $p$ with galactocentric radius
to present and future observational data.

Another possible extension of the model is to incorporate a gaseous halo (corona) and turbulent pumping (turbulent diamagnetism).
One effect of this would be the ability of the dynamo to withstand larger outflow speeds \citep{Gressel10}.
The effect of a gaseous halo, including turbulent diffusivity $\beta$ that varies with height,
has not been studied in the context of dynamical quenching theory.

In certain ways, our model may be oversimplified. 
For example, the turbulent correlation time $\tau\corr$ may not, in fact,
be equal to the eddy turnover time $\tau$ for supernova-driven turbulence \citep{Shukurov04}.
Moreover, the parameters $H$, $\tau$, and $V$ may vary significantly with position from the midplane $z$.
It also deserves to be mentioned that turbulent transport coefficients such as $\alpha$ (or in general $\alpha_{ij}$)
are known to experience rotational quenching and to become more anisotropic as $\Omega\tau$ increases beyond $\sim1$
\citep[][and references therein]{Ruzmaikin+88,Brandenburg+Subramanian05a}.
As expressions in our model are, in any case, imprecise to within factors of order unity,
and as we have restricted our analysis to values of $\Omega\tau\le2$,
the neglect of these effects is not an important limitation of the present study.
However, possible modifications to the model for large $\Omega\tau$ are worth exploring in a future study.

Introducing non-axisymmetry into the dynamo equations may result in a number of effects, 
including new contributions to the $\Emf$ term \citep{Brandenburg+Subramanian05a}.
In particular, spiral shocks would tend to align magnetic field along the spiral arms \citep{Vaneck+15},
and generally streaming flows associated with spiral density waves might lead to important effects on $\meanv{B}$.
It should be noted, however, that at least for the most extreme case of M33, 
alignment of magnetic field with spiral arms would not help because the pitch angle of the spiral arms is smaller in magnitude
than that inferred for the mean magnetic field of that galaxy \citep[][and references therein]{Vaneck+15}.

One still highly contentious area of mean-field dynamo theory is the nature of the magnetic helicity flux 
and the corresponding flux density term in the dynamical quenching Equation~\eqref{dalpha_mdt} for $\alpha\magn$.
As we have shown, the value of $p$ is independent of the nonlinearity used 
\textit{for a certain class} of nonlinearities \citep{Chamandy+14b}.
However, 
it is possible that our model excludes important flux terms 
\citep{Vishniac+Cho01,Subramanian+Brandenburg06,Vishniac12b,Vishniac+Shapovalov14,Ebrahimi+Bhattacharjee14},
that may drastically affect the mean magnetic field and its pitch angle.

In fact, we are aware of a few alternate models that potentially lead to large pitch angles.
A mean-field dynamo for which the $\alpha$ term is dominated by the Vishniac-Cho (VC) flux 
leads to $p\sim-\tan^{-1}[\pi/(2H)]\sim-17^\circ$ for $H=5$ \citep{Sur+07}.
However, it is not clear how such a dynamo would saturate.
The dynamo model of \citet{Moss+99} based on the buoyancy-driven Parker instability 
results in estimates of the pitch angle that can be very large (their Equation~(19)).
This model is worthy of more exploration in the future.
Notably, both the VC flux and Parker instability mechanisms exhibit threshold behavior,
in that they require a mean magnetic field of a certain strength to be present initially.

Another promising avenue for producing large pitch angles is the magnetorotational instability (MRI) 
\citep{Kitchatinov+Rudiger04,Elstner+09}.
The unstable MRI modes have $p=-45^\circ$ in the ideal MHD limit,
but including finite viscosity and diffusivity (e.g. due to turbulence) 
might result in considerably smaller $|p|$ \citep{Pessah+Chan08}.
In any case, the influence of the MRI on the galactic magnetic field deserves further study.

\section{Summary and conclusions}
\label{sec:conclusions}
Magnitudes of magnetic pitch angles of the large-scale magnetic fields of spiral galaxies 
are generally underestimated by nonlinear dynamo theory.
For the 31 data points compiled by \citet{Vaneck+15}, $p$ has a mean value of $-25^\circ$ 
and minimum and maximum values of $-8^\circ$ and $-48^\circ$, excluding uncertainties,
whereas standard nonlinear dynamo theory with canonical parameters generally predicts $|p|<15^\circ$.

We have taken a fresh look at the predictions of $p$ from standard mean-field galactic dynamo theory
without making any \textit{a priori} judgements about parameter values.
Our basic model is a thin disk (slab) dynamo subject to magnetic helicity balance, 
that grows the mean magnetic field to a steady saturated state.
We have shown that in this model solutions for $p$ in the saturated state (or in the kinematic regime for that matter)
can be usefully parameterized in terms 
of the angular velocity $\Omega$, shear parameter $q\equiv-\del\ln\Omega/\del\ln r$, scale height of the disk
in turbulent correlations lengths $H\equiv h/l$, turnover time of energy-carrying eddies $\tau\equiv l/u$,
and mean outflow velocity in units of the rms turbulent velocity of energy-carrying eddies $V\equiv U\f/u$.
Of these, $\Omega$ and $q$ are well-constrained observationally, leaving a 3-dimensional parameter space,
which is dimensionless if the inverse Rossby number $\Omega\tau$ is used in place of $\tau$.
The range of allowed parameter space is further restricted by theoretical considerations,
namely the requirements that the mean magnetic field grows rather than decays 
and that the kinetic energy in the helical part of the turbulence is less than or equal to the total turbulent kinetic energy.

This still leaves the standard parameter space as well as a rather small non-standard region of parameter space 
that leads to $|p|\gtrsim20^\circ$, and a much smaller region that gives $|p|\gtrsim40^\circ$.
To be specific, such large $|p|$ requires smaller than canonical $H$ and larger than canonical $\tau$,
which can be achieved simultaneously and rather naturally 
by adopting a turbulent correlation scale $l$ that is a few times its canonical value of $0.1\kpc$,
if the standard value $h=0.5\kpc$ is assumed.
Larger than canonical values of $V$ are also required for the largest $|p|$;
on the other hand, dynamos with larger $l$ have larger kinematic growth rates 
and are better able to withstand a large $V$ without becoming subcritical.
We have argued, based on independent evidence from the literature,
that such changes to the parameter values are plausible. 
However, the need to adopt parameters near one extreme of the allowed parameter space
suggests that physical elements may be missing from this basic mean-field dynamo model,
and we have offered several suggestions for extending the model to include potentially important effects.

Our results were obtained using asymptotic analytical as well as numerical solutions,
and are almost independent of the precise form of the dynamo nonlinearity assumed.
Good agreement was obtained between the two types of solution for most of the parameter space.
The analytical solution becomes inaccurate for parameters for which the $\alpha^2$ effect becomes important.
Interestingly, these are the parameters that lead to the largest values of $|p|$,
rendering numerical solutions indispensable.

A more detailed comparison between theory and observation is warranted.
A likelihood analysis of the parameter space, given the \citet{Vaneck+15} data, 
would be interesting both in its own right and as a demonstration
of what could be done with better data sets and perhaps better models.
A related goal would be model comparison between the basic model presented and models that incorporate new physical effects,
but as a result may require more free parameters.

\section*{Acknowledgements}
We thank Anvar Shukurov and Andrew Fletcher for valuable discussions
and for helpful comments on an early draft of the manuscript,
and David Moss for useful feedback on a later version.
We are also grateful to Romeel Dav\'{e}, Oliver Gressel, Marita Krause, Jonathan Zwart, Nathan Deg, Prasun Dutta, Pallavi Bhat,
and especially Kandaswamy Subramanian for useful discussions.
Finally, we thank the anonymous referee for insightful suggestions that helped to improve the paper.
\bibliographystyle{apj}
\bibliography{refs}
\end{document}